\documentclass[12pt]{article}
\usepackage{amsmath, amsthm, amssymb,slashed}
\usepackage{hyperref}
\newcommand{\be}{\begin{equation}}
\newcommand{\ee}{\end{equation}}
\newcommand{\bea}{\begin{eqnarray}}
\newcommand{\eea}{\end{eqnarray}}

\begin{document}
\begin{titlepage}
\begin{center}
{\bf\Large{\vbox{\centerline{On neutral scalar radiation by  a massive orbiting star in }\vskip .5cm
 \centerline{ extremal Kerr-Newman black hole}}}}
\vskip
0.5cm
{Nan Bai$^{a}$ \footnote{bainan@itp.ac.cn}, Yi-Hong Gao$^{a}$ \footnote{gaoyh@itp.ac.cn}, Xiao-bao Xu$^{a}$ \footnote{xbxu@itp.ac.cn}} \vskip 0.05in
{\it ${}^a$ State Key Laboratory of Theoretical Physics,\\
Institute of Theoretical Physics,\\ Chinese Academy of
Sciences, P.O. Box 2735, Beijing 100190, China }

\end{center}
\vskip 0.5in
\baselineskip 16pt
\abstract{In this  short note we extend the work of 1401.3746 about gravity waves by a massive orbiting star in an extremal Kerr black hole to an extremal Kerr-Newman black hole for scalar radiation, and still find that  it has a CFT interpretation from   Kerr-Newman/CFT. In addition, we investigate on electromagnetic radiation with Kerr/CFT, which a detailed analysis isn't given by 1401.3746.}
\end{titlepage}

\section{Introduction}
Detecting gravitational waves is an important work, recently BICEP2\cite{bicep2} found primordial gravitational waves which provide a more firmer evidence on inflation. Black hole radiation is also the source of gravitational waves, recently \cite{porstr}\cite{porstr1} make use of conformal symmetry studying gravitational radiation produced by a massive orbiting star very near the horizon of an extreme Kerr black hole, they find analytic results and provide a new evidence on  Kerr/CFT conjecture\cite{kerrcft}, which is that bulk quantum gravity on the NHEK(Near-Horizon Extreme Kerr) geometry of Kerr is dual to a two-dimensional boundary CFT. The orbiting star sits at $r_0$  along the radial direction of Kerr black hole, where the boundary at $r\to \infty$.  The orbiting star is a perturbative source and  breaks the scaling symmetry of the bulk theory, however if we assume that there is a renormalization group  flow between UV CFT and IR CFT on the boundary, then according to holographic renormalization group flow\cite{holrg}\cite{holrg1}\cite{holrg2}, so the bulk theory of $r<r_0$ corresponds to IR CFT. This is what \cite{porstr} did,  then the gravitational radiation follows from the application of  Fermi¡¯s golden rule to the  IR CFT, they get a nontrivial agreement between bulk and boundary calculations.

One simple extension of Kerr/CFT is Kerr-Newman/CFT \cite{kncft}, in \cite{kncft1} the authors calculate near-superradiant scattering of charged scalars and fermions in a near-extreme Kerr-Newman black hole and find that Kerr-Newman/CFT is also hold in this case. Now we want to generalize the consideration in\cite{porstr}  simply to Kerr-Newman case and make a small check on Kerr-Newman/CFT conjecture. For simplicity, we also consider a massless neutral scalar field radiation, then we find that the bulk gravitational radiation is also agreement with the CFT result.

For electromagnetic and gravitational perturbations in Kerr black hole, Teukolsky \cite{Teukolsky}\cite{Teukolsky1}\cite{Teukolsky2}\cite{Teukolsky3} find using Newman-Penrose formalism \cite{Chandra}, the equation of perturbation is decoupled into angular part and radial part. For the NHEK geometry, the similar analysis is done in \cite{0906.2380}\cite{0906.2376}, we will use their result to analyze the electromagnetic perturbation along with 1401.3746.

To make the note more readable, in section 2 we briefly review Kerr-Newman black hole, NHEKN and  NHEKN/CFT. In section 3 we calculate the massless neutral scalar radiation both from the bulk gravity theory and the boundary CFT. In section 4 we give the calculation of electromagnetic perturbation with Kerr/CFT. The last section is some discussion.

\section{Review Kerr-Newman black hole, NHEKN and  NHEKN/CFT}
The metric of the Kerr-Newman black hole with mass $M$, angular momentum $J=aM$, and electric charge $Q$ in the Boyer-Lindquist coordinates is ($G=\hbar=c=1$)\cite{kncft1}:
\be\label{KerrNewman}
ds^2=-{\Delta \over\hat \rho^2}\left(d\hat t-a \sin^2\theta d\hat\phi\right)^2+{\sin^2 \theta \over  \hat{\rho}^2}
\left((\hat r^2+a^2)d\hat \phi-a d\hat t\right)^2+{\hat{\rho}^2 \over\Delta}d\hat r^2+ \hat{\rho}^2 d\theta^2\,
\ee
where
\be
\Delta=\hat r^2+a^2-2M\hat{r}+Q^2\,,\quad \hat {\rho}^2=\hat r^2 +a^2\cos^2 \theta
\ee
And the gauge field and field strength are
\begin{align}
A &= -\frac{Q\hat{r}}{\hat{\rho}^2}(d\hat{t}-a\sin^2\theta d\hat{\phi}) \,\\
F &= -\frac{Q(\hat{r}^2-a^2\cos^2\theta)}{\hat{\rho}^4}(d\hat{t}-a\sin^2\theta d\hat{\phi})\wedge d\hat{r}\cr
  &\quad -\frac{2Q\hat{r}a\cos\theta}{\hat{\rho}^4}\sin\theta d\theta\wedge(a d\hat{t}-(\hat{r}^2+a^2)d\hat{\phi})\nonumber
\end{align}
The horizon $r_{\pm}$ are
\be
r_\pm = M \pm \sqrt{M^2 - a^2 - Q^2}
\ee
The entropy,  Hawking temperature, angular velocity of the horizon, and electric potential are
\bea
S &=& {\mbox{Area}\over 4} = \pi (r_+^2 + a^2) \\
T_H  &=& {r_+ - r_-\over 4\pi(r_+^2+ a^2)} \notag\\
\Omega_H &=& \frac{ a}{r_+^2 + a^2}  \notag\\
\Phi &=& {Q r_+ \over r_+^2 + a^2} \,\notag
\eea
We also define the dimensionless Hawking temperature
\be
\tau_H \equiv {r_+ - r_-\over r_+}
\ee

Now we consider near-horizon extremal Kerr-Newman geometry. The extremal black hole has $r_+=r_-=M$, then we  take near horizon limit by defining\cite{barhoro}\cite{kncft}
\begin{align}\label{NearHor}
&\hat{r}=r_+ + \lambda r_0 r\ ,\cr
&\hat{t}=t\,r_0/\lambda\ ,\cr
&\hat{\phi}=\phi+\Omega_H\frac{t\, r_0}{\lambda}
\end{align}
with $r_0^2=r_++a^2$. Taking $\lambda\to0$, the near horizon geometry is
\be\label{NHEKN}
 ds^2=\Gamma(\theta)\left[
-r^2dt^2+\frac{dr^2}{r^2}
+ d\theta^2 \right] + \gamma(\theta)(d\phi+b\, rdt)^2
\ee
where
\bea\label{definefuncs}
\Gamma(\theta) &=& r_+^2 + a^2 \cos^2\theta \cr
\gamma(\theta) &=& \frac{(r_+^2+a^2)^2\sin^2\theta}{r_+^2 +a^2 \cos^2\theta} \cr
b &=& {2ar_+\over r_+^2 + a^2}
\eea

NHEKN has an  enhanced isometry group $U(1)_L \times SL(2,R)_R$ , generated by the Killing vectors
\begin{align}
Q_0 &= \partial_\phi\ ,\cr
H_1 &= \partial_t\ ,\cr
H_2 &= t \partial_t - r \partial_r\, \cr
H_3 &= \left({1\over 2r^2} + {t^2\over 2}\right)\partial_t - t r \partial_r - {b\over r}\partial_ \phi\
\end{align}

The first law of thermodynamics for Kerr-Newman is
\be
T_H dS = dM - \Omega_H dJ - \Phi dQ \,\label{thermo}
\ee
At extremality, $T_H=0$,  the above formula reduces to $dM = \Omega_H dJ + \Phi dQ$, the first law becomes
\be\label{extremalfirstlaw}
dS = {1\over T_L}\left(dJ - \mu_L dQ\right) \,
\ee
where\cite{porstr}
\bea\label{knpotentials}
T_L &=& {r_+^2 + a^2\over 4 \pi J}\\
\mu_L &=& - {Q^3 \over 2 J} \ \notag
\eea
According to the Kerr/CFT correspondence, $T_L$ is the left moving temperature  of the dual 2D CFT.

\section{Scalar radiation from a star orbiting near the horizon}
In this section we calculate a massless neutral scalar radiation  caused by a massive orbiting star in the NHEKN geometry from the bulk gravity, and compare with the result of the  boundary CFT. We find that both calculations agree with each other well, so this supports the Kerr-Newman/CFT conjecture.
\subsection{Gravity analysis in NHEKN}
There is a  star orbiting at radius $r_0$ in the NHEKN geometry \eqref{NHEKN}. We  parameterize the corresponding geodesic $x_*^\mu(t)$ with the NHEKN time $t$:
\bea\label{plunge in NHEKN}
x_*^t(t)&=&t\,,\cr
x_*^\phi(t)&=&\phi_0-\omega_s r_0\,t\,, \cr
x_*^r(t)&=&r_0\,,\cr
x_*^\theta(t)&=&{\pi \over 2}\,.
\eea
where the velocity $\omega_s$  of the orbiting star
\be
\omega_s=\frac{(4a^2-M^2)M}{2a(M^2+a^2)}
\ee
The orbiting star is moving along its geodesic in the NHEKN geometry,  its energy is zero
\be
E=-g_{t\mu}\partial_\tau x_*^\mu=0
\ee
where $\tau$ is the proper time.
The angular momentum (per unit mass) is
\be
L=g_{\phi \mu}\partial_\tau x_*^\mu=\frac{M^2+a^2}{\sqrt{4a^2-M^2}}
\ee
We couple the star to a massless neutral scalar with the interaction
\be\label{Sint}
S_I=4\pi\lambda \int d\tau \, \Psi(x_*(\tau))\,
\ee
where  $\lambda$ is a coupling constant.
Then the scalar wave equation in the presence of the star is:
\be
\square\, \Psi=4\pi {\cal T}\,
\ee
where (setting  $\phi_0=0$)
\begin{align}\label{T for NHEKN plunge}
{\cal T}&=-\lambda(-g)^{-{1\over2}}\frac{d\tau}{dt}\delta(r-r_0)\delta(\theta-{\pi\over 2})\delta(\phi+\omega_s r_0 t)\cr
       &=-\lambda\frac{M\sqrt{4a^2-M^2}}{2a}r_0\Gamma^3(\theta)\gamma(\theta)\bigg |_{\theta={\pi\over2}}\delta(r-r_0)\delta(\theta-{\pi\over 2})\delta(\phi+\omega_s r_0 t)\cr
       &=-\lambda\frac{\sqrt{4a^2-M^2}}{2aM(M^2+a^2)}r_0\delta(r-r_0)\delta(\theta-{\pi\over 2})\delta(\phi+\omega_s r_0 t)
\end{align}
where $g$ is the determinant of the metric.
The orbiting star preserve the Killing symmetry:
\be
\chi=\partial_t-\omega_s r_0 \partial_\phi
\ee
Then a $\chi$-invariant solution to the wave equation is
\begin{eqnarray}
\Psi&=&\sum_{\ell,m}e^{im\left(\phi+\omega_s r_0 t\right)}S_\ell(\theta) R_{\ell m}(r)\,,\label{psi simpler  expansion}\\
\,{\cal T}&=&{1 \over 4 \pi (M^2+a^2\cos^2\theta)}\sum_{\ell,m}e^{im\left(\phi+\omega_s r_0 t\right)}S_\ell(\theta) T_{\ell m}(r)\,\label{source simpler expansion}
\end{eqnarray}
where $S_\ell$ are the spheroidal harmonics obeying
\be {1\over\sin\theta}\partial_\theta(\sin\theta\,\partial_\theta S_\ell)+\left(K_\ell-\frac{m^2}{\sin^2\theta}-\frac{a^4\, m^2}{(r_+^2+a^2)^2}\sin^2\theta\right)S_\ell= 0\,\label{angular eqn}
\ee
with $K_l$ is a separation constant, $\ell\ge 0\,, -\ell\le m\le \ell$. Note that $S_\ell$ and $K_\ell$ depend on both $m$ and $\ell$.
The spheroidal harmonics satisfy the normalization condition:
\be\label{sph harmonics normn}
\int_0^\pi\sin\theta d\theta\,S_\ell(\theta) S_{\ell'}(\theta)=\delta_{\ell \ell'}\,
\ee
The coefficients of ${\cal T}$ are
\be
T_{\ell m}(r)
=-\lambda\frac{M\sqrt{4a^2-M^2}}{a(M^2+a^2)}r_0 S_\ell(\pi/2)\delta(r-r_0)\, \label{T_lm simple}
\ee
The separated radial equation is
\be
\partial_r(r^{2}\partial_r R_{\ell m})+\left(\frac{\omega^2}{r^2}+c_2\frac{2\omega m}{r}+\frac{(2ar_+)^2}{(r_+^2+a^2)^2}m^2+\frac{2a^2}{r_+^2+a^2}m^2-K_l\right)R_{\ell m}=T_{\ell m}\,\label{radial eqn}
\ee
where $c_2=\frac{2ar_+}{r_+^2+a^2}$  and $\omega=-m\omega_sr_0$

There are two linear independent solutions to the homogeneous radial equation ($T_{\ell m}=0$), they are the Whittaker functions\cite{0906.2376}:
\be\label{Whittaker solns}
M_{imc_2,h-{1\over2}}\left(-{2i\omega}/{r}\right)\,,\quad W_{imc_2,h-{1\over2}}\left(-{2i\omega}/{r}\right)\,
\ee
where
\be\label{h defn}
h\equiv{1\over2}+\sqrt{1/4+K_\ell-\frac{2a^2(3r_+^2+a^2)}{(r_+^2+a^2)^2}m^2}\,
\ee
We also consider the case $h$ is real and $h<1$  as in \cite{porstr}.

We then find the solution of \eqref{radial eqn} that is purely ingoing at the horizon and obeys Dirichlet boundary condition at $r=\infty$, see the similar process in \cite{porstr}
\be\label{radial soln}
R_{\ell m}(r)={1\over W}\left[X\,\Theta(r_0-r) W_{imc_2,h-{1\over2}}\left(-{2i\omega}/{r}\right) +Z\,\Theta(r-r_0) M_{imc_2,h-{1\over2}}\left(-{2i\omega}/{r}\right)\right]
\ee
The Wronskian W of the two linear independent solution \eqref{Whittaker solns},
\be
W=2i\omega\frac{\Gamma(2h)}{\Gamma(h-imc_2)}
\ee
and
\begin{align}\label{coeffsolu}
X=-\lambda\frac{M\sqrt{4a^2-M^2}}{a(M^2+a^2)}r_0 S_\ell(\pi/2)  M_{imc_2,h-{1\over2}}\left(2i\omega_s\,m\right)\,\cr
Z=-\lambda\frac{M\sqrt{4a^2-M^2}}{a(M^2+a^2)}r_0 S_\ell(\pi/2)  W_{imc_2,h-{1\over2}}\left(2i\omega_s\,m\right)\,
\end{align}
So it is easy shown that the Klein-Gordon particle number flux
\be
{\cal F}=-\int\sqrt{-g}J^r d\theta d\phi\,,\quad J^\mu={i\over 8\pi}(\Psi^*\nabla^\mu\Psi-\Psi\nabla^\mu\Psi^*)\,,\quad \nabla_\mu J^\mu=\mathrm{Im}[\Psi {\cal T}]\,
\ee
is vanishing at the infinity for real h, while at the horizon we get
\bea\label{particle number flux}
{\cal F}_{\ell m}&=&-(r_+^2+a^2)\frac{\omega}{2}\,\bigg |\frac{X_{lm}}{W_{lm}}\bigg|^2 e^{-\pi mc_2}\cr
&=&\frac{\lambda^2M^2(4a^2-M^2)r_0}{8m\omega_sa^2(M^2+a^2)}S_\ell^2(\pi/2)\bigg | M_{imc_2,h-{1\over2}}\left(2i\omega_s\,m\right)\bigg|^2\cr
\qquad &&\bigg |\frac{\Gamma(h-imc_2)}{\Gamma(2h)}\bigg|^2 e^{-\pi mc_2}
\eea
The above formula gives the particle number flux across the future horizon.

\subsection{CFT analysis}
In this section we calculate the particle number flux  from the boundary CFT in the spirit of \cite{kncft1}, follows from  AdS/CFT\cite{hep-th/9808017}\cite{hep-th/9812007}, the massless neutral scalar field in the bullk deforms the boundary CFT to
\be\label{pt}
S = S_{CFT}+\sum_{\ell}\int dt^+dt^-J_{\ell}(t^+,t^-)\mathcal{O}_{\ell}(t^+,t^-)\,.
\ee
It was shown that \cite{porstr}
\be
J_\ell(\phi,t)=\sum_m\frac{X}{W}C\,e^{im(\phi+\omega_s r_0t^-)}
\ee
where we identify that
\be
t^+=\phi\quad t^-=t\quad  C=(-2i\omega)^{1-h}\frac{\Gamma(2h-1)}{\Gamma(h-imc_2)}
\ee
And the operator $\mathcal{O}$ dual to the scalar field has conformal weight \cite{bhss}:
\be
h_L=h_R=h
\ee
where $h$ is equal to \eqref{h defn}. Then by using Fermi¡¯s golden rule, the transition rate \cite{hep-th/9702015}\cite{hep-th/9706100}
\be\label{total rate}
\mathcal{R}=2\pi \sum_{\ell, m}|J_{\ell m}|^2 \int dt^+ dt^-  e^{-imt^+- im3r_0t^-/4}G(t^+,t^-)\,.
\ee
where $G(t^+,t^-)=\langle \mathcal{O}^\dagger(t^+,t^-)\mathcal{O}(0,0)\rangle_{T_L}$ is the two point function of the dual two-dimensional conformal field theory, its specific form is\cite{kncft1}
\be
G=C_{\mathcal{O}}^2(-1)^{h_L+h_R}\bigg(\frac{\pi T_L}{\sinh(\pi T_L t^+)}\bigg)^{2h_L}\bigg(\frac{\pi T_R}{\sinh(\pi T_R t^-)}\bigg)^{2h_R}e^{iq_L\mu_Lt^++iq_R\mu_Rt^-}
\ee
where $C_{\mathcal{O}}$ is a normalization constant.

So from Kerr-Newman/CFT, we have $q_L=e,~\mu_L=-\frac{Q^3}{2J},~T_L=\frac{M^2+a^2}{4\pi M a}$, there is
\bea\label{parate}
R_{\ell m}&=&C_{\mathcal{O}}^22\pi\frac{|X_{\ell m}|^2}{(2h-1)^2}(4\omega^2)^{-h}\frac{(2\pi T_L)^{2h-1}}{\Gamma(2h)^2}e^{-\frac{m}{2T_L}}|\Gamma(h+i\frac{m}{2\pi T_L})|^2\cr
&&(2\pi T_R)^{2h-1}e^{\frac{m\omega_s r_0-q_R \mu_R}{2T_R}}|\Gamma(h+i\frac{m\omega_s r_0-q_R \mu_R}{2\pi T_R})|^2\cr
&=&C_{\mathcal{O}}^2\frac{2\pi}{(2h-1)^2}\frac{\lambda^2 M^2(4a^2-M^2)}{a^2 (M^2+a^2)^2}r_0^2 S_{\ell}^2(\pi/2)|M_{imc_2,h-\frac{1}{2}}(2i\omega_s m)|^2\frac{(2\pi T_L)^{2h-1}}{\Gamma(2h)^2(4\omega_s^2m^2r_0^2)^h}e^{-\frac{m}{2T_L}}\cr
&&|\Gamma(h+i\frac{m}{2\pi T_L})|^2(2\pi T_R)^{2h-1}e^{\frac{m\omega_s r_0-q_R \mu_R}{2T_R}}|\Gamma(h+i\frac{m\omega_s r_0-q_R \mu_R}{2\pi T_R})|^2\cr
&=&C_{\mathcal{O}}^2\frac{2\pi}{(2h-1)^2}\frac{\lambda^2 M^2(4a^2-M^2)}{a^2 (M^2+a^2)^2}r_0^2 S_{\ell}^2(\pi/2)|M_{imc_2,h-\frac{1}{2}}(2i\omega_s m)|^2\frac{(2\pi T_L)^{2h-1}}{\Gamma(2h)^2(4\omega_s^2m^2r_0^2)^h}e^{-\pi m c_2}\cr
&&|\Gamma(h+imc_2)|^2(2\pi T_R)^{2h-1}e^{\frac{m\omega_s r_0-q_R \mu_R}{2T_R}}|\Gamma(h+i\frac{m\omega_s r_0-q_R \mu_R}{2\pi T_R})|^2
\eea
where we have set $q_L=0$, because  we consider a neutral scalar.

And in the limit $T_R\to0$, we have
\bea\label{parate1}
R_{\ell m}&=&C_{\mathcal{O}}^2 \frac{2\pi}{(2h-1)^2}\frac{\lambda^2 M^2(4a^2-M^2)}{a^2 (M^2+a^2)^2}r_0^2 S_{\ell}^2(\pi/2)|M_{imc_2,h-\frac{1}{2}}(2i\omega_s m)|^2\frac{(2\pi T_L)^{2h-1}}{\Gamma(2h)^2(4\omega_s^2m^2r_0^2)^h}e^{-\pi m c_2}\cr
&&|\Gamma(h+imc_2)|^2 p_R^{2h-1}2\pi
\eea
where we have define $p_R=m\omega_s r_0-q_R\mu_R$.
If we assume $q_R\mu_R\ll1$ as \cite{porstr}, the above formula becomes to
\bea\label{cftresult}
R_{\ell m}&=& C_{\mathcal{O}}^2\frac{(2\pi)^2}{(2h-1)^2}\frac{\lambda^2 M^2(4a^2-M^2)}{a^2 (M^2+a^2)^2}r_0^2 S_{\ell}^2(\pi/2)|M_{imc_2,h-\frac{1}{2}}(2i\omega_s m)|^2\frac{1}{2^{2h}\omega_s m r_0} (\frac{M^2+a^2}{2Ma})^{2h-1}e^{-\pi m c_2}\cr
&&\frac{|\Gamma(h+imc_2)|^2}{\Gamma(2h)^2}\cr
&=&C_{\mathcal{O}}^2 \frac{(2\pi)^2}{(2h-1)^2}\frac{1}{2^{2h}}(\frac{M^2+a^2}{2Ma})^{2h-1}\frac{\lambda^2 M^2(4a^2-M^2)r_0} {a^2 (M^2+a^2)^2 \omega_s m}e^{-\pi m c_2}S_{\ell}^2(\pi/2)|M_{imc_2,h-\frac{1}{2}}(2i\omega_s m)|^2\cr
&&\frac{|\Gamma(h+imc_2)|^2}{\Gamma(2h)^2}
\eea

\subsection{Comparison}
If we take
\be
C_{\mathcal{O}}^2=\frac{2^{2h}(2h-1)^2}{(2\pi)^2}(\frac{2Ma}{M^2+a^2})^{2h-1}\frac{1}{8}(M^2+a^2)
\ee
the particle flux number in both calculations are exactly equal, i.e. ${\cal F}_{\ell m}=R_{\ell m}$.
And when the charge $Q$ of Kerr-Newman black hole is vanishing, we can also recover the result in \cite{porstr}.

\section{Electromagnetic radiation in Kerr/CFT}
Using null tetrad $(l,n,m,\bar m)$, the scalar of electromagnetic field
\be
\phi_2=F_{\mu\nu}\bar{m}^{\mu}n^{\nu}
\ee
obeys the following perturbation equation
\be\label{Teukolsky phi2 eqn}
[(\Delta+\gamma-\bar{\gamma}+2\mu+\bar{\mu})(D+2\epsilon-\rho)-(\bar{\delta}+\alpha+\bar{\beta}+2\pi-\bar{\tau})(\delta-\tau+2\beta)]\delta\phi_2=-2\pi J_2
\ee
where
\be\label{J2}
J_2=(\Delta+\gamma-\bar{\gamma}+2\mu+\bar{\mu})J_{\bar{m}}-(\bar{\delta}+\alpha+\bar{\beta}+2\pi-\bar{\tau})J_{n}
\ee
The electromagnetic current of the orbiting star with charge $q$ is
\be
J^{\nu}=q\int d\tau(-g)^{-1/2}\frac{dx^{\nu}}{d\tau}\delta^{(4)}(x^\alpha-x^\alpha_*(\tau))
\ee
For NHEK, the Kinnersley tetrad and nonzero spin coefficients can be seen in \cite{porstr}.
The nonzero component of the current along the geodesic
\begin{eqnarray}\label{geodesic in NHEK}
x_*^t(t)&=&t\,,\cr
x_*^\phi(t)&=&-\frac{3}{4}r_0\,t\,, \cr
x_*^r(t)&=&r_0\,,\cr
x_*^\theta(t)&=&{\pi \over 2}\,.
\end{eqnarray}
is
\be
J_\phi=\frac{r_0 q}{2M^2}\delta^{(4)}(x^\alpha-x^\alpha_*(\tau))
\ee
From \eqref{J2} we have
\begin{eqnarray}\label{J2plunge}
J_2&=&\frac{i r_0^2 q}{8\sqrt{2}M^5}\Big[4\delta(r-r_0)\delta(\theta-\pi/2)\delta(\phi+3r_0 t/4) \nonumber\\&&\qquad +r_0\delta'(r-r_0)\delta(\theta-\pi/2)\delta(\phi+3r_0t/4)\nonumber\\&&\qquad-\frac{3}{4}\delta(r-r_0)\delta(\theta-\pi/2)\delta'(\phi+3r_0t/4)\nonumber\\&&\qquad
-2i\delta(r-r_0)\delta'(\theta-\pi/2)\delta(\phi+3r_0t/4)\Big]
\end{eqnarray}
The equation \eqref{Teukolsky phi2 eqn}is separable for the variable
\be
\Psi^{(-1)}\equiv\eta^{-2}\delta\phi_2
\ee
For $s=-1$, the perturbation equation (2.5) of \cite{0906.2380} is
\begin{align}
&\frac{1}{r^2}\partial^2_t\Psi^{(-1)}-\frac{2}{r}\partial_t\partial_\phi\Psi^{(-1)}+(1-\frac{(1+\cos^2\theta)^2}{4\sin^2\theta})\partial^2_\phi\Psi^{(-1)}\cr
&-r^2\partial^2_r\Psi^{(-1)}-\frac{1}{\sin\theta}\partial_\theta(\sin\theta\partial_\theta\Psi^{(-1)})+\frac{2}{r}\partial_t\Psi^{(-1)}\cr
&+2(i\frac{\cos\theta}{\sin^2\theta}+i\frac{\cos\theta}{2})\partial_\phi\Psi^{(-1)}+(\cot^2\theta+1)\Psi^{(-1)}=-{\mathcal J}_2
\end{align}
where ${\mathcal J}_2=2\pi 2M^2\eta^{-2} J_2/(\eta\bar{\eta})$.
According to the symmetry of the background geometry, we use the following expansion
\begin{eqnarray}
\Psi^{(-1)}&=&\sum_{\ell,m}e^{im\left(\phi+3r_0t/4\right)}S_\ell(\theta) R_{\ell m}(r)\,,\label{psi simpler  expansion s}\\
2\pi{\cal T}&=&\eta\bar\eta \sum_{\ell,m}e^{im\left(\phi+3r_0t/4\right)}S_\ell(\theta) T_{\ell m }(r)\,,\label{source simpler expansion s}
\end{eqnarray}
where $\mathcal T=2M^2\eta^{-2} J_2$ and $S_\ell$ are the spin-weighted spheroidal harmonics obeying
\be
{1\over\sin\theta}\partial_\theta(\sin\theta\,\partial_\theta S_\ell)+\left(K_\ell-\frac{m^2+1-2m\cos\theta}{\sin^2\theta}- \frac{m^2}{4}\sin^2\theta+m\cos\theta\right)S_\ell=0\,,\label{angular eqn s=-1}
\ee
The radial equation is
\be
r^2\partial^2_r R_{\ell m}+(2m^2-K_{\ell}+\frac{2\omega(m+i)}{r}+\frac{\omega^2}{r^2})R_{\ell m}=T_{\ell m} \label{radialelectromagnetic}
\ee
where
\be\label{Tfourier}
T_{\ell m}=\frac{i r_0^2q}{4\sqrt{2}M^3}[(2iS'_\ell(\frac{\pi}{2})-\frac{3}{4}i m S_\ell(\frac{\pi}{2}))\delta(r-r_0)+r_0S_l(\frac{\pi}{2})\delta'(r-r_0)]
\ee
From the result of gravitational perturbation \cite{porstr}, the solution of radial equation \eqref{radialelectromagnetic} satisfying the physical boundary condition is
\be\label{radial soln s=-1}
R_{\ell m}(r)={1\over r_0^2 W} \left[\mathcal{X}\,\Theta(r_0-r) \mathcal{W}(r)+ \mathcal{Z}\, \Theta(r-r_0) \mathcal{M}(r)\right],
\ee
where $W=2i\omega\frac{\Gamma(2h)}{\Gamma(h-im+1)}$ and
\begin{eqnarray}
\mathcal{X}&=& r_0\mathcal{M}'(r_0)(-a_1)+\mathcal{M}(r_0)(a_0+2a_1)\cr
\mathcal{Z}&=&\mathcal{X}(\mathcal{M}\to\mathcal{W})
\end{eqnarray}
And $a_0$ and $a_1$ is followed from \eqref{Tfourier}, i.e.
\begin{eqnarray}
a_0&=&\frac{i r_0^2q}{4\sqrt{2}M^3}(-\frac{3imS}{4}+2iS')\cr
a_1&=&\frac{i r_0^2q}{4\sqrt{2}M^3}S
\end{eqnarray}

The asymptotic behavior of the solution is immediately from \cite{porstr}
\begin{eqnarray}
\Psi^{(-1)}(r\to 0)&=&\sum_{\ell,m} e^{im(\phi+3r_0t/4)}\,S_\ell(\theta)\,\frac{\mathcal{X}}{r_0^{2}W} (-2i\omega)^{im-1}\, r^{-im+2}e^{-3imr_0/4r}\,,\\
\Psi^{(-1)}(r\to \infty)&=&\sum_{\ell,m}e^{im(\phi+3r_0t/4)}\,S_\ell(\theta)\,\frac{\mathcal{Z}}{r_0^{2}W} (-2i\omega)^{h} \,r^{-h+1}\,.
\end{eqnarray}
where $h\equiv\frac{1}{2}+\sqrt{1/4+K_\ell-2m^2}$ and we assume $h$ is real.
The photon number flux at the horizon is given by \cite{Teukolsky3}
\bea\label{photonnumber}
\mathcal{F}_{\ell m}&=&\frac{\frac{3}{4}r_0 M^6}{|\mathcal B|^2}m e^{-\pi m}\left|\frac{\mathcal{X}}{r_0^2 W}\right|^2\cr
|\mathcal B|^2&\equiv &(K_\ell-m^2)^2+m^2
\eea

\subsection{CFT analysis}
The source term in the dual CFT can be read from the Herz potential of electromagnetic perturbation, which is given by
\be
\Psi_H=\frac{1}{\mathcal B}R^{(-1)}(r)S^{(+1)}(\theta)e^{-i\omega t+i m\phi}
\ee
where $\omega=-\frac{3}{4}m r_0$. And
\bea
&&R^{(-1)}(r)=\frac{\mathcal{X}}{r_0^{2} W} \mathcal{W}(r)\\
&&~~~~\to \frac{\mathcal{X}}{ r_0^{2} W} \left[\frac{(-2i\omega)^{1-h}\Gamma(2h-1)}{\Gamma(h+1-im)}r^{h}+ \frac{(-2i\omega)^h\Gamma(1-2h)}{\Gamma(2-h-im)}r^{-h+1}\right]\quad \textrm{for $r\to \infty$}\,.\notag
\eea
So the source term is
\be
J_{\ell m}=\frac{1}{\mathcal {B}}\frac{\mathcal{X}}{r_0^{2} W}\frac{(-2i\omega)^{1-h}\Gamma(2h-1)}{\Gamma(h+1-i m)}
\ee
Then the transition rate is
\be
R_{\ell m}=2\pi |J_{\ell m}|^2 C^2_{\mathcal O}\frac{1}{\Gamma(2h_L)}e^{-\pi m}|\Gamma(h_L+i m)|^2\frac{2\pi}{\Gamma(2h_R)}(\frac{3}{4}m r_0)^{2h_R-1}
\ee
According to Kerr/CFT \cite{kncft1}, we have
\be
h_R=h\,,\quad h_L=h+1\,.
\ee
If we choose the normalization factor
\be
C^2_{\cal O}=\frac{2^{2(h-1)}M^6}{(2\pi)^2}\frac{\Gamma(2h+2)\Gamma(2h)}{\Gamma(2h-1)^2}\,,
\ee
we can get the bulk result \eqref{photonnumber}.

\section{Conclusion}
In this short note, we calculate the particle flux number of a massless neutral scalar produced by a orbiting star in the NHEKN geometry from the bulk gravity and the boundary CFT and find the agreement between them. We also investigate the electromagnetic radiation along with \cite{porstr}. This is a little trivial case, it would be interesting to consider the electrical charged scalar waves or gravitational waves in the NHEKN geometry, and compare with calculations of the CFT.

\section*{Acknowledgements}
We thank  A. P. Porfyriadis for very useful correspondence.


\begin{thebibliography}{13}

\bibitem{bicep2}  P. A. R. Ade et al. [BICEP2 Collaboration], ``BICEP2 I: Detection Of B-mode Polarization at Degree
Angular Scales", arXiv:1403.3985 [astro-ph.CO].
\bibitem{porstr} A. P. Porfyriadis, A. Strominger, ``Gravity Waves from Kerr/CFT'', arXiv:1401.3746.
\bibitem{porstr1} S. Hadar, A. P. Porfyriadis, A. Strominger, ``Gravity Waves from Extreme-Mass-Ratio Plunges into Kerr Black Holes'', arXiv:1403.2797.
\bibitem{kerrcft} M. Guica, T. Hartman, W. Song and A. Strominger, ``The Kerr/CFT Correspondence'',
Phys. Rev. D 80, 124008 (2009). [arXiv:0809.4266].
\bibitem{holrg} D. Z. Freedman, S. S. Gubser, K. Pilch and N. Warner, ``Renormalization Group Flows
from Holography, Supersymmetry, and a c-Theorem'', Adv. Theor. Math. Phys. 3, 363  1999. [arXiv:hep-th/9904017].
\bibitem{holrg1} I. R. Klebanov, A. A. Tseytlin, ``Asymptotic Freedom and Infrared Behavior in the Type 0 String Approach to Gauge Theory'',
Nucl. Phys B 547,143  1999. [arXiv:hep-th/9812089].
\bibitem{holrg2} P. Kraus, F. Larsen and S. Trivedi, ``The Coulomb Branch of Gauge Theory from
Rotating Branes'',  JHEP 9903,  003 (1999). [arXiv:hep-th/9811120].
\bibitem{kncft} T. Hartman, K. Murata, T. Nishioka and A. Strominger, ``CFT Duals for Extreme
Black Holes'', JHEP 0904, 019 (2009). [arXiv:0811.4393].
\bibitem{kncft1} T. Hartman, W. Song and A. Strominger, ``Holographic Derivation of Kerr-Newman
Scattering Amplitudes for General Charge and Spin'', JHEP 1003, 118 (2010). [arXiv:0908.3909].
\bibitem{barhoro} J. M. Bardeen and G. T. Horowitz, ``The extreme Kerr throat geometry: A vacuum
analog of AdS(2) x S(2)'', Phys. Rev. D 60, 104030 (1999). [arXiv:hep-th/9905099].
\bibitem{bhss} I. Bredberg, T. Hartman, W. Song and A. Strominger, ¡°Black Hole Superradiance From
Kerr/CFT,¡± JHEP 1004, 019 (2010) [arXiv:0907.3477 [hep-th]].
\bibitem{0906.2376}  A. J. Amsel, G. T. Horowitz, D. Marolf and M. M. Roberts,
  ``No Dynamics in the Extremal Kerr Throat'',JHEP 0909, 044 (2009).
  [arXiv:0906.2376].
\bibitem{hep-th/9808017}
  V. Balasubramanian, P. Kraus, A. E. Lawrence and S. P. Trivedi,
  ``Holographic probes of anti-de Sitter space-times,''
  Phys. Rev.  D  59, 104021 (1999).
  [arXiv:hep-th/9808017].
\bibitem{hep-th/9812007}
  U. H. Danielsson, E. Keski-Vakkuri and M. Kruczenski,
  ``Vacua, propagators, and holographic probes in AdS / CFT,''
  JHEP  9901, 002 (1999).
  [arXiv:hep-th/9812007].
\bibitem{hep-th/9702015} J. M. Maldacena and A. Strominger, ``Universal low-energy dynamics for rotating black
holes'', Phys. Rev. D 56, 4975 (1997). [arXiv:hep-th/9702015].
\bibitem{hep-th/9706100}  S. S. Gubser, ``Absorption of photons and fermions by black holes in four-dimensions'',
Phys. Rev. D 56, 7854 (1997). [arXiv:hep-th/9706100].
\bibitem{Teukolsky} S. A. Teukolsky,
  ``Rotating black holes - separable wave equations for gravitational and electromagnetic perturbations'', Phys. Rev. Lett. 29, 1114 (1972).
\bibitem{Teukolsky1} S. A. Teukolsky,
``Perturbations of a rotating black hole. 1. Fundamental equations for gravitational electromagnetic and neutrino field perturbations'',  Astrophys. J. 185, 635 (1973).
\bibitem{Teukolsky2}  W. H. Press and S. A. Teukolsky,
  ``Perturbations of a Rotating Black Hole. II. Dynamical Stability of the Kerr Metric'',
  Astrophys. J. 185, 649 (1973).
\bibitem{Teukolsky3}  S. A. Teukolsky and W. H. Press, ``Perturbations of a rotating black hole. III - Interaction of the hole with gravitational and electromagnet ic radiation'', Astrophys. J. 193, 443 (1974).
\bibitem{Chandra} S. Chandrasekhar,
  ``The Mathematical Theory of Black Holes'', (Oxford University Press, New York, 1983).
\bibitem{0906.2380}
  O. J. C. Dias, H. S. Reall and J. E. Santos,
  ``Kerr-CFT and gravitational perturbations'',
  JHEP 0908, 101 (2009). [arXiv:0906.2380].










\end{thebibliography}
\end{document}